\documentclass[reqno,11pt]{amsart}
\usepackage{epsfig}\usepackage{amsfonts,amsmath,amssymb,mathrsfs,verbatim,amsthm,amscd}
\newcommand{\unit}{1\!\!1}

\newcommand{\beq}{\begin{equation}}
\newcommand{\ee}{\end{equation}}
\newcommand{\bea}{\begin{eqnarray}}
\newcommand{\eea}{\end{eqnarray}}

\usepackage{hyperref}

\newcommand{\p}{\partial_x}
\newcommand\id{1\kern-0.25em\text{l}}
\newcommand\zero{0\kern-0.4em\text{0}}

\newcommand{\R}{\mathbb R}
\newcommand{\Z}{\mathbb Z}

\newcommand{\be}{\begin{equation}}
\newcommand{\ba}{\begin{eqnarray}}
\newcommand{\ea}{\end{eqnarray}}

\begin{document}

\title[Weak supersymmetry and superconformal indices]
{Weak supersymmetry and superconformal indices}

\author{Vyacheslav \,P. Spiridonov}%

\address{Laboratory of Theoretical Physics,
JINR, Dubna, Moscow region, 141980 Russia and
National Research University Higher School of Economics, Moscow, Russia
}

\maketitle

\vspace*{-2em}

\begin{abstract}
One-dimensional quantum mechanical models obeying Smilga's weak supersymmetry are described in the matrix form.
They are related to the parasupersymmetric and higher-order derivative deformations
of the standard supersymmetric models discussed earlier.
Their superconformal indices are computed and shown to be equivalent
to the Witten indices for the second order supersymmetric quantum mechanics.
\end{abstract}

\begin{flushright}
\em To the memory of Valery Rubakov
\end{flushright}

\vspace*{1em}

The notion of superconformal indices was introduced in \cite{Rom1,KMMR}. It represents an
analogue of the Witten index \cite{Wi81} for supersymmetric systems obeying a non-standard
superalgebra of symmetries. Witten indices count the number of vacua of the systems (the
states with the minimal energy), whereas superconformal indices deal with the BPS
states defined as zero modes of the supercharges (for a review, see \cite{RR2016}).

Explicit computation of the superconformal indices in four-dimensional $\mathcal{N}=1$ supersymmetric
gauge field theories  in \cite{DO} revealed that they are expressed in terms of elliptic
hypergeometric integrals---a novel class of special functions discovered just
eight years before \cite{spi:umn}. R\"omelsberger conjectured \cite{Rom2}
that superconformal indices of the models related by the Seiberg electromagnetic duality \cite{Sei1,Sei2}
coincide. This conjecture was proven by Dolan and Osborn in \cite{DO} on the basis of rigorous
mathematical theorems, the simplest of which was established in  \cite{spi:umn}
and confirmed the duality considered in \cite{Sei1} in the sector
of BPS states. Exact computability of the univariate elliptic beta integral
serves as an indicator on the $s$-confinement phenomenon, when the magnetic dual theory
has not local gauge symmetry. For the original general Seiberg duality \cite{Sei2}, the equality of
superconformal indices for electric and magnetic theories requires a more complicated
theorem established in \cite{Rains}. A very large number of mathematical relations
following from known conjectural Seiberg type dualities was described in \cite{SV,SV2}.
In the opposite direction, many new novel dual field theories were suggested on the basis of known
mathematical identities for the elliptic hypergeometric integrals. For a review of the theory of
elliptic hypergeometric functions, see \cite{rev}.

In the present paper we discuss one-dimensional quantum mechanical models obeying weak
supersymmetry --- a deformation of supersymmetry suggested by Smilga in \cite{Sweak}.
It is based on the centrally extended superalgebra $su(2|1)$, which emerged earlier in the
supersymmetric field theories quantized on the space-time $\mathrm{R}\times\mathrm{S}^3$
\cite{Sen}. The coincidence of algebras emerging in \cite{Sweak} and \cite{Sen} is established in \cite{Sind}
(a quantum mechanical model with a similar symmetry algebra was considered also in \cite{BN}).
Here, we reconsider these models in the line of earlier deformations (``weakenings'')
of the standard supersymmetric quantum mechanics (SQM),
which were described in old works \cite{RS,AISV,SMPL92,AIS}.

The minimal $\mathcal{N}=1$ SQM models is based on the $su(1|1)$ superalgebra having the form
\begin{equation}\label{susy}
\{Q^+,Q^-\}=H, \qquad [H,Q^\pm]=(Q^\pm)^2=0,
\end{equation}
where $H$ is the Hamiltonian, which can be considered as a central element of the algebra,
and $Q^\pm$ are conserved supercharges. For simplification,
we assume that $H$ and all other Hamiltonians to be used below are defined
by self-adjoint operators having only discrete spectra.
The operators $Q^\pm$ and their analogues defined below are supposed to be hermitian conjugates
of each other, $(Q^\pm)^\dag =Q^\mp$. In terms of the self-adjoint operators
$Q_1=Q^++Q^-, Q_2=(Q^+-Q^-)/\textup{i}$ the algebra
looks like $\{Q_j,Q_k\}=2H\delta_{jk},\; [H,Q_j]=0,\, j, k=1,2.$

Two evident consequences of these relations are: 1) the eigenvalues $\lambda_a$ of the Hamiltonian,
$H|\psi^{(a)}\rangle=\lambda_a|\psi^{(a)}\rangle$, satisfy the inequality
$\lambda_a\geq0$ and 2) all energy levels with $\lambda_a>0$ are
at least doubly degenerate. Supersymmetry is not spontaneously broken if there exists a lowest
energy state, a vacuum, which is a zero mode of supercharges, $Q^{\pm}|0\rangle=0$.
Clearly, if it exists, it must have zero energy, $H|0\rangle=0$.
So, the Hamiltonian spectrum consists of an infinite direct sum of two-dimensional
irreducible representations of the algebra \eqref{susy} fixed by positive eigenvalues
of $H$ and possible one-dimensional representations fixed by the zero modes of $H$.

The Witten index introduced in \cite{Wi81} has the form
\begin{equation}\label{WI}
I_W=Tr\,(-1)^Fe^{-\beta H}, \quad \beta>0,
\end{equation}
where $F$ is the fermion charge and the trace is taken over the complete set of orthonormal
eigenfunctions, $Tr\,A=\sum_{a} \langle\psi^{(a)}|A|\psi^{(a)}\rangle$. This index counts
the difference of the numbers of bosonic and fermionic vacua, $I_W=n_b-n_f$.
If $I_W \neq 0$, the supersymmetry is not broken.

For an explicit realization of relations \eqref{susy} and their analogues to be considered below
we use physical models of a spin 1/2 (or spin 1) particle on the line $x\in\R$ in the
potential $U(x)$ and an external magnetic field $B(x)$ along the vertical direction.
The corresponding Hamiltonian has the form
$H_{phys}=\frac{1}{2m} \hat p^2+U(x)+B(x)J_3$, $\hat p=-\textup{i}\hbar d/dx$. For simplification 
of formulas we impose
the normalization of the mass $m=1$ and the Planck constant $\hbar=1$. Usually the operator
$H$ in \eqref{susy} is rewritten as $2H_{phys}$, which we do not do because we wish to
remove the coefficient $\tfrac12$ from many formulas.

The simplest realization of the algebra \eqref{susy} emerges for a special relation between
the potentials $U(x)$ and $B(x)$. Namely, from the supercharges
\begin{equation}
Q^+=\left(\begin{array}{cc}
 0 & A^+ \\
 0   & 0
\end{array}\right),\quad
Q^-=\left(\begin{array}{cc}
 0 & 0 \\
 A^-   & 0
\end{array}\right),
\quad A^{\pm}=\mp \p +v(x),\quad \p:=\frac{d}{dx},
\label{SUSYchar}\end{equation}
one gets the Hamiltonian
\begin{equation}
H=\left(\begin{array}{cc}
h_1 & 0 \\
 0   &  h_2
\end{array}\right)=-\p^2+v^2(x)-v'(x)\sigma_3,
\quad \sigma_3=\left(\begin{array}{cc}
1 & 0\\
 0   & -1
\end{array}\right),
\label{HamSUSY}\end{equation}
where $v'(x)\equiv \p v(x)$.

Clearly, this construction is deeply related to the factorization method of solving the Schr\"odinger
equation initiated by Schr\"odinger himself, see survey \cite{IH}. This method uses an infinite
chain of Hamiltonians, all of which are factorized up to some constants as products of differential operators
of the first order conjugated to each other
\beq
L_j=A_j^+A_j^-+\lambda_j, \qquad A_j^{\pm}=\mp \p +v_j(x), \quad j=0,\pm 1, \pm2, \ldots
\label{fact}\ee
The neighboring Hamiltonians are obtained by permutation of the $A_j^\pm$ operators,
i.e. one demands validity of the following factorization chain
\beq
 L_{j+1}=A_j^-A_j^+ +\lambda_j=A_{j+1}^+A_{j+1}^-+\lambda_{j+1}.
\label{fchain}\ee
Then each pair $L_j-\lambda_j, L_{j+1}-\lambda_j$ can be used as the supersymmetric subhamitonians.
Supercharges \eqref{SUSYchar} are realized as products of the bosonic $A_j$ and fermionic $f^\pm$ operators,
$Q^+=A_j^+f^+,\, Q^-=A_j^-f^-$, where the matrices
$$\displaystyle
f^+=\left(\begin{array}{cc}
 0 & 1\\
 0   & 0
\end{array}\right),\quad
f^-=\left(\begin{array}{cc}
 0 & 0 \\
 1   & 0
\end{array}\right)
$$
satisfy the fermionic creation and annihilation operator algebra
$$(
f^-)^2=(f^+)^2=0, \quad  \{f^-,f^+\}=1.
$$

As mentioned, we assume that $L_j$ are self-adjoint $L_j^\dag=L_j$, and $A_j^\pm=(A_j^\mp)^\dag.$
Then
$$
\int_\R\psi^*(x)L_{j+1}\psi(x)dx=||A_j^+\psi||^2+\lambda_j=||A_{j+1}^-\psi||^2+\lambda_{j+1}
$$
for any square integrable function $\psi(x)\in\mathrm{L}^2(\R)$ of the unit norm.
Therefore, all eigenvalues of $L_j$ operators $L_j\psi^{(j)}(x) =E^{(j)}\psi^{(j)}(x)$
satisfy the constraints $E^{(j)}\geq\lambda_{j-1},\, \lambda_j$.

As a result, one has the intertwining relation $L_{j+1}A_j^-=A_j^-L_j$ (and its conjugate
$A_j^+L_{j+1}=L_j A_j^+$) meaning that the
eigenfunctions of the neighboring Hamiltonians are related as $\psi^{(j+1)}(x)=A_j^-\psi^{(j)}(x)$
together with an appropriate connection between the norms of these eigenfunctions. Such a relation
between solutions of two differential equations of the second order is called the Darboux
transformation in the theory of integrable systems.
All this leads to the relations
\beq
L_j=-\partial_x^2 +u_j(x),\quad
u_j(x)=v_j^2(x)-v_j'(x)+\lambda_j, \quad u_{j+1}(x)=u_j(x)+2v_j'(x),
\label{ujfj}\ee
and the key equation
\beq
 v_j^2(x)-v_{j+1}^2(x)+v_j'(x)+v_{j+1}'(x)=\lambda_{j+1}-\lambda_j.
\label{chain}\ee
This chain was considered for the first time as a functional differential-difference equation of two continuous
variables $x$ and $j$ by Infeld in \cite{Inf}.

In \cite{RS}, Rubakov and the author suggested a deformation of SQM by replacing the fermion by a parafermion of order $p=2$ whose creation and annihilation operators satisfy the relations
$$
a^3=0, \quad a^2a^+ + a^+a^2= 2a,\quad  aa^+a =2a
$$
and their partners following from the hermitian conjugation rules $a^\dag=a^+,\, (a^+)^\dag =a$.
Using natural realization of $a^+, a$ operators by $3\times 3$ matrices, the following generalization of
supercharges was suggested in \cite{RS}
$$
Q^+=\left(\begin{array}{ccc}
 0 & A_1^+ & 0 \\
 0   & 0 & A_2^+ \\
 0 & 0& 0
\end{array}\right),\quad
Q^-=\left(\begin{array}{ccc}
 0 & 0 & 0 \\
 A_1^-  & 0 & 0 \\
 0 & A_2^- & 0
\end{array}\right), \quad (Q^\pm)^3=0,
$$
which generated the following parasuperalgebra
\begin{eqnarray}\nonumber &&
(Q^-)^2Q^+ + Q^-Q^+Q^- +Q^+(Q^-)^2=2Q^-H, \quad [H, Q^-]=0,
\\  &&
(Q^+)^2Q^- + Q^+Q^-Q^+ +Q^-(Q^+)^2=2Q^+H, \quad [H, Q^+]=0,
\label{PSUSYalg1}\end{eqnarray}
where the Hamiltonian $H$ is given by a diagonal matrix of the form
$$
H=-\p^2 +\textrm{diag}(v_1^2-v_1'-c, v_1^2+v_1' -c , v_2^2+v_2'+c)
$$
with an arbitrary real constant $c$.
In terms of the described above Hamitonians $L_j$
$$
H=\textup{diag} (h_1, h_2, h_3), \quad h_j=L_j-\kappa,\quad \kappa =\frac{\lambda_1+\lambda_2}{2}.
$$
The subamiltonian $h_2$ can be written in the form $-\p^2+v_2^2-v_2'+c$
because the functions $v_1(x)$ and $v_2(x)$ satisfy the equation
\begin{equation}
v_1'(x)+v_2'(x)+v_1^2(x)-v_2^2(x)=\lambda_2-\lambda_1 \equiv 2c.
\label{cons}\end{equation}
The superpotentials $W_j(x)$ and the constant $c$ used in \cite{RS} correspond to our $-v_j(x)$
and $-2c$, respectively. Below we assume that $c\neq 0$, the choice $c=0$ is trivial for our purposes.
The case $c<0$ can be reached from $c>0$ by the changes $v_{1,2}(x)\to -v_{2,1}(x)$.
After some additional constraints on superpotentials, the Hamiltonian of the parasupersymmetric quantum
mechanical (PSQM) model takes the form of the Hamiltonian for a spin 1 particle on a line with a magnetic field
along the vertical axis \cite{RS}. Note that by construction the spectrum of general
Hamiltonian $H$ is triply degenerate with possible exception of two lowest levels.

It was shown in \cite{AISV} that this PSQM system can be obtained from the standard algebra
\eqref{susy} after applying the ansatz \eqref{SUSYchar} with the superpotential $v(x)$ being
given by the $2\times 2$ matrix function. With the help of additional constraints,
the resulting supersymmetric Hamiltonian can be
given a block-diagonal form and the PSQM Hamiltonian corresponds to its $3\times3$
submatrix. This means that PSQM represents a truncated version of SQM. However, the
constraints on the system now become weaker and even the negative eigenvalues
of the Hamiltonian become admissible when full supersymmetric algebra cannot
be realized by self-adjoint operators.

Factorization \eqref{fact} is not unique. One can take an invertible operator $T$
and replace $A^+_j$ and $A_j^-$ by $\tilde A_j^+=A^+_jT$ and $\tilde A_j^-=T^{-1}A_j^-$, respectively.
This preserves the rules of hermitian conjugation, if $T$ is a unitary operator, $T^\dag=T^{-1}$.
In general, the operator $T^{-1}A_j^-A_j^+T$, the supersymmetric partner of $\tilde A_j^+\tilde A_j^-$,
is not obliged to have the Schr\"odinger form. However, for the unitary affine transformation,
$Tf(x)=\sqrt{q}f(qx+a)$, the operator
$$
q^{-2}\tilde A_j^-\tilde A_j^+=-\p^2+q^{-2}\left(v_j^2(\tfrac{x-a}{q})-v_j'(\tfrac{x-a}{q})\right)
$$
is a Schr\"odinger operator again and one can define a $q$-deformed supersymmetric quantum mechanics (qSQM)
based on the algebra
$$
Q^+Q^-+q^{-2}Q^-Q^+=H,\quad (Q^\pm)^2=0, \quad HQ^\pm=q^{\pm2}Q^\pm H,
$$
where $Q^\pm$ have the same form as before with $A_j^\pm$ replaced by $\tilde A_j^\pm$.
For a special self-similar potential showing exponential discrete spectrum one has both
the $q$-deformed algebra and the double degeneracy away from the ground state  \cite{SMPL92}.

Another version of ``smashing'' the standard SQM was proposed in the work \cite{AIS}.
It is obtained from the same ansatz for supercharges as in \eqref{SUSYchar} after
replacing $A_j^\pm$ by the higher order differential operators keeping the
physical Hamiltonian in the standard Schr\"odinger form. Because of that it was
called the higher-order (or higher-derivative) supersymmetric quantum mechanics (HSQM).
In this paper we use only the second order case, when supercharges $Q^\pm$ have the following form
\begin{equation}
Q^+=\left(\begin{array}{cc}
 0 & A_1^+A_2^+ \\
 0   & 0
\end{array}\right), \quad
Q^-=\left(\begin{array}{cc}
 0 & 0 \\
A_2^-A_1^-  & 0
\end{array}\right).
\label{HSUSYch}\end{equation}
Their anticommutator produces a quadratic polynomial of the Hamiltonian $H$
$$
\{Q^+, Q^-\}= (H-c)(H+c), \qquad [H, Q^\pm]=(Q^\pm)^2=0, \quad 2c=\lambda_2-\lambda_1,
$$
where
\begin{equation}
H=\left(\begin{array}{cc}
-\p^2+v_1^2-v_1'-c & 0  \\
 0 & -\p^2+v_2^2+v_2'+c
\end{array}\right) =
\left(\begin{array}{cc}
L_1-\kappa & 0  \\
 0 & L_3-\kappa
\end{array}\right).
\label{HamHSQM}\end{equation}
A direct connection to PSQM models mentioned above is evident---one just deletes in
the corresponding $3\times 3$ diagonal matrix Hamiltonian the middle term.

If one replaces in the ansatz \eqref{HSUSYch} the product $A_1^+A_2^+$ by $A_1^+\cdots A_n^+T$
and $A_2^-A_1^-$ by $T^{-1}A_n^-\cdots A_1^-$, where $T$ is the affine transformation operator,
then one obtains a general $q$-deformed polynomial superalgebra
$$
Q^+Q^-+q^{-2}Q^-Q^+=\prod_{k=1}^n (H-\lambda_k), \quad (Q^\pm)^2=0,\quad
HQ^\pm=q^{\pm 2}Q^\pm H, \quad
$$
where
$$
H=\left(\begin{array}{cc}
h_1 & 0  \\
 0 & h_2
\end{array}\right),\quad h_1=A_1^+A_1^-+\lambda_1,\quad h_2= q^{-2}T^{-1}A_n^-A_n^+T+\lambda_n.
$$
Imposing a natural physical constraint that this $H$ describes a spin 1/2 particle in the homogeneous
magnetic field along the vertical axis, one comes to a very rich class of self-similar potentials
(in some sense the most general known class of exactly solvable potentials of quantum mechanics).
In our $n=2$, $q=1,\, a=0$ case the Hamiltonian is shifted by a constant $c$ for matching with the PSQM
Hamiltonian. It will be important also for the weak supersymmetry model to be discussed below.
For a detailed description of HSQM models and general self-similar
potentials, see \cite{AI} and \cite{S95}, respectively.

Finally, we discuss the weak supersymmetric quantum mechanical (WSQM) model,
which was introduced by Smilga in \cite{Sweak}. This is one more deformation
of the standard SQM, additional to PSQM, HSQM, qSQM cases mentioned before.
Consider the following $4\times 4$ matrix supercharges
$$
Q_1^-=\left(\begin{array}{cccc}
 0 & 0 & 0 & 0\\
A_1^- & 0 & 0 & 0 \\
A_1^- & 0 & 0 & 0 \\
0 & A_2^- & -A_2^- & 0 \\
\end{array}\right),
\quad
Q_2^-=\left(\begin{array}{cccc}
 0 & 0 & 0 & 0\\
-A_1^- & 0 & 0 & 0 \\
A_1^- & 0 & 0 & 0 \\
0 & A_2^- & A_2^- & 0 \\
\end{array}\right)
$$
and their hermitian conjugates $Q_\alpha^+=(Q_\alpha^-)^\dag$,
$$
Q_1^+=\left(\begin{array}{cccc}
 0 & A_1^+ & A_1^+ & 0\\
 0 & 0 & 0 & A_2^+ \\
 0 & 0 & 0 & -A_2^+ \\
0 & 0 & 0 & 0 \\
\end{array}\right),
\quad
Q_2^+=\left(\begin{array}{cccc}
 0 & -A_1^+ & A_1^+ & 0\\
 0 & 0 & 0 & A_2^+  \\
 0 & 0 & 0 & A_2^+ \\
0 & 0 & 0 & 0 \\
\end{array}\right).
$$
They generate now the $su(2|1)$ superalgebra with a central extension
\begin{equation}
\{Q_\alpha^\pm,Q_\beta^\pm\}=0,\quad \{Q_\alpha^-,Q_\beta^+\}=2\big((H-Y)\delta_{\alpha\beta}+Z_{\alpha\beta}\big),
\label{WSQM}\end{equation}
where $H=\textup{diag}(h_1,h_2,h_3,h_4)$ is the Hamiltonian having the diagonal form
\begin{equation}
H=-\p^2+\textup{diag} (v_1^2-v_1'-c,v_1^2+v_1'-c,v_2^2-v_2'+c,v_2^2+v_2'+c).
\label{ham_wsqm}\end{equation}
It differs from the Hamiltonian of PSQM model just by the doubling of the middle subhamiltonian,
i.e. $h_2=h_3$. Other operator entries in the relation \eqref{WSQM} are
\begin{eqnarray*} &&  \makebox[0.7em]{}
Y=c\left(\begin{array}{cccc}
 -1 & 0 & 0 & 0\\
0 & 0 & 0 & 0 \\
0 & 0 & 0 & 0 \\
0 & 0 & 0 & 1 \\
\end{array}\right),
\quad
Z_{11}=c\left(\begin{array}{cccc}
0 & 0 & 0 & 0\\
0 & 0 & 1 & 0 \\
0 & 1 & 0 & 0 \\
0 & 0 & 0 & 0 \\
\end{array}\right)=-Z_{22},
 \\  &&
Z_{12}=c\left(\begin{array}{cccc}
0 & 0 & 0 & 0\\
0 & -1 & 1 & 0 \\
0 & -1 & 1 & 0 \\
0 & 0 & 0 & 0 \\
\end{array}\right),
\quad Z_{21}=Z_{12}^\dag=
c\left(\begin{array}{cccc}
0 & 0 & 0 & 0\\
0 & -1 & -1 & 0 \\
0 & 1 & 1 & 0 \\
0 & 0 & 0 & 0 \\
\end{array}\right).
\end{eqnarray*}

All the described charges commute with the Hamiltonian, which thus plays the role of the central charge,
$$
[H,Q_\alpha^\pm]=[H,Y]=[H,Z_{\alpha\beta}]=0.
$$
Other commutation relations of the algebra look as follows
\begin{equation}
\quad [Q_\alpha^\pm, Y]=\pm c Q_\alpha^\pm,\quad [Y,Z_{\alpha\beta}]=0.
\label{comm1}\end{equation}
Denoting
$$
J_0:=\tfrac{1}{c}Z_{11},\quad J_+:=\tfrac{1}{2c}Z_{12}\quad J_-:=\tfrac{1}{2c}Z_{21},
$$
one comes to the $sl(2)$ algebra commutation relations
$$
[J_0, J_\pm]=\pm 2J_\pm,\quad [J_+,J_-]=J_0
$$
together with the complimentary identities
\begin{equation}
[Q_\alpha^\pm, J_0]=\pm Q_\alpha^\pm,\quad [Q_1^\pm, J_\pm]=\pm Q_2^\pm,
\quad [ Q_2^\mp, J_\pm]=\mp Q_1^\mp,
\label{comm2}\end{equation}
and vanishing commutators
$$
[Q_1^\mp, J_\pm]=[Q_2^\pm, J_\pm]=0.
$$

The described WSQM model was originally formulated by Smilga \cite{Sweak} using the creation and annihilation
operators for two fermions. Below we establish exact correspondence between such a formulation with
our $4\times4$ matrix model. We remark that there exists also a superfield formulation of this
system given in \cite{IS}. Let us introduce the following $4\times 4$ matrices
$$
f_1^+=f^+\otimes\unit=
\left(\begin{array}{cccc}
0 & 0 & 1 & 0 \\
0 & 0 & 0 & 1 \\
0 & 0 & 0 & 0 \\
0 & 0 & 0 & 0 \\
\end{array}\right), \quad
f_1^-=f^-\otimes\unit=
\left(\begin{array}{cccc}
0 & 0 & 0 & 0 \\
0 & 0 & 0 & 0 \\
1 & 0 & 0 & 0 \\
0 & 1 & 0 & 0 \\
\end{array}\right),
$$
$$
f_2^+=(-1)^{f_1^+f_1^-}\unit\otimes f^+=
\left(\begin{array}{cccc}
0 & -1 & 0 & 0 \\
0 & 0 & 0 & 0 \\
0 & 0 & 0 & 1 \\
0 & 0 & 0 & 0 \\
\end{array}\right), \quad
f_2^-=(-1)^{f_1^+f_1^-}\unit\otimes f^-=
\left(\begin{array}{cccc}
0 & 0 & 0 & 0 \\
-1 & 0 & 0 & 0 \\
0 & 0 & 0 & 0 \\
0 & 0 & 1 & 0 \\
\end{array}\right),
$$
where $\unit$ is the $2\times2$ unit matrix.
They define the needed two-fermion algebra
$$
\{f_\alpha^\pm,f_\beta^\pm\}=0,\qquad \{f_\alpha^-,f_\beta^+\}=\delta_{\alpha\beta},\quad  \alpha, \beta=1,2.
$$
Apply to them the following unitary transformation
$$
\psi_\alpha^\pm=Uf_\alpha^\pm U^{-1}, \quad U= \frac{1}{\sqrt{2}}\left(\begin{array}{cccc}
 1 & 0 & 0 & 1 \\
0 & 1 & 1 & 0 \\
0 & -1 & 1 & 0 \\
-1 & 0 & 0 & 1 \\
\end{array}\right),  \quad U^{-1}= \frac{1}{\sqrt{2}}\left(\begin{array}{cccc}
 1 & 0 & 0 & -1 \\
0 & 1 & -1 & 0 \\
0 & 1 & 1 & 0 \\
1 & 0 & 0 & 1 \\
\end{array}\right)
$$
preserving the fermionic anticommutation relations
$$
\{\psi_\alpha^\pm,\psi_\beta^\pm\}=0,\qquad \{\psi_\alpha^-,\psi_\beta^+\}=\delta_{\alpha\beta},\quad  \alpha, \beta=1,2.
$$
As a result, we obtain
$$
\psi_1^-=\frac{1}{2}\left(\begin{array}{cccc}
0 & 1 & -1 & 0 \\
1 & 0 & 0 & -1 \\
1 & 0 & 0 & -1 \\
0 & 1 & -1 & 0 \\
\end{array}\right), \qquad
\psi_2^-=\frac{1}{2}\left(\begin{array}{cccc}
0 & 1 & 1 & 0 \\
-1 & 0 & 0 & 1 \\
1 & 0 & 0 & -1 \\
0 & 1 & 1 & 0 \\
\end{array}\right),
$$
as well as
$$
\psi_1^+=\frac{1}{2}\left(\begin{array}{cccc}
0 & 1 & 1 & 0 \\
1 & 0 & 0 & 1 \\
-1 & 0 & 0 & -1 \\
0 & -1 & -1 & 0 \\
\end{array}\right), \qquad
\psi_2^+=\frac{1}{2}\left(\begin{array}{cccc}
0 & -1 & 1 & 0 \\
1 & 0 & 0 & 1 \\
1 & 0 & 0 & 1 \\
0 & 1 & -1 & 0 \\
\end{array}\right).
$$

Now we can rewrite all the entries in the above algebra $Q_\alpha^\pm, H, Y, J_0, J_\pm$
in terms of these fermion operators.  Direct computations yield
$$
[\psi_1^+,\psi_1^-]=\left(\begin{array}{cccc}
0 & 0 & 0 & -1 \\
0 & 0 & -1 & 0 \\
0 & -1 & 0 & 0 \\
-1 & 0 & 0 & 0 \\
\end{array}\right), \quad
[\psi_2^+,\psi_2^-]=\left(\begin{array}{cccc}
0 & 0 & 0 & -1 \\
0 & 0 & 1 & 0 \\
0 & 1 & 0 & 0 \\
-1 & 0 & 0 & 0 \\
\end{array}\right),
$$
$$
\psi_2^+[\psi_1^+,\psi_1^-]=\frac{1}{2}\left(\begin{array}{cccc}
0 & -1 & 1 & 0 \\
-1 & 0 & 0 & -1 \\
-1 & 0 & 0 & -1 \\
0 & 1 & -1 & 0 \\
\end{array}\right), \quad
\psi_1^-[\psi_2^+,\psi_2^-]=\frac{1}{2}\left(\begin{array}{cccc}
0 & -1 & 1 & 0 \\
1 & 0 &  & -1 \\
1 & 0 & 0 & -1 \\
0 & -1 & 1 & 0 \\
\end{array}\right).
$$

Let us denote $v_1(x)=v(x)+B(x),\, v_2(x)=v(x)-B(x)$.
Then the constraint \eqref{cons} can be easily resolved, which yields
\begin{equation}
B(x)=\frac{c-v'(x)}{2v(x)}
\label{B}\end{equation}
for an arbitrary function $v(x)$. In \cite{RS} the opposite choice was proposed---to keep
$B(x)$ as an arbitrary function and to determine $v(x)$ in terms of $B(x)$ as
\begin{equation}
v(x)=c g(x)^{-1}\int_{x_0}^x g (y)dy,\quad g(x)=\exp\Big(\int^x 2B(y)dy)\Big),\quad c\neq 0,
\label{g}\end{equation}
i.e. it is $g(x)$ that can be taken as an arbitrary free function with $2B(x)=g'(x)/g(x)$.
As a result, one can write
$$
A_2^\pm+B(x) =A_1^\pm -B(x)=\mp\p+v(x).
$$
Note that our $B(x)$ differs by the sign from the one introduced in \cite{Sweak}.

Then,
$$
Q_1^-=(\p+v(x))(\psi_1^-+\psi_2^+)+B(x)(\psi_1^-[\psi_2^+,\psi_2^-]-\psi_2^+[\psi_1^+,\psi_1^-]),
$$
$$
Q_2^-=(\p+v(x))(\psi_2^- -\psi_1^+)+B(x)(\psi_1^+[\psi_2^+,\psi_2^-] + \psi_2^-[\psi_1^+,\psi_1^-])
$$
and
$$
Q_1^+=(-\p+v(x))(\psi_1^+ + \psi_2^-)+B(x)(\psi_1^+[\psi_2^+,\psi_2^-]-\psi_2^-[\psi_1^+,\psi_1^-]),
$$
$$
Q_2^+=(-\p+v(x))(\psi_2^+ -\psi_1^-)+B(x)(\psi_1^-[\psi_2^+,\psi_2^-] + \psi_2^+[\psi_1^+,\psi_1^-]).
$$

Other charges can be represented in the following bilinear form
$$
Y=c(\psi_1^-\psi_2^- + \psi_2^+\psi_1^+), \quad
J_0=\psi_1^-\psi_1^+ -\psi_2^-\psi_2^+,\quad
J_+=\psi_1^-\psi_2^+, \quad J_-=\psi_2^-\psi_1^+,
$$
whereas the Hamiltonian is quartic in the fermionic operators
\begin{eqnarray}\nonumber &&
H=-\p^2+f^2(x)+B^2(x) +2 v'(x)(\psi_1^-\psi_2^- + \psi_2^+\psi_1^+)
\\  && \makebox[3em]{}
- B'(x) [\psi_1^+,\psi_1^-][\psi_2^+,\psi_2^-].
\label{Hamiferm}\end{eqnarray}

Now it is necessary to match with the notation used in the papers \cite{Sweak,Sind}.
Denote $\psi_\alpha:=\psi_\alpha^-$
and $\psi^\alpha=\epsilon^{\alpha\beta}\psi_\beta$, where $\epsilon^{\alpha\beta}=-\epsilon^{\beta\alpha}
=-\epsilon_{\alpha\beta},\, \epsilon^{12}=1$. Set also $\bar\psi^\alpha =\psi_\alpha^\dag=\psi_\alpha^+$.
In this notation
$Q_\alpha:=Q_\alpha^-,\; \bar Q^\alpha:=Q_\alpha^+$ and the algebra is rewritten as
\begin{equation}
\{Q_\alpha,Q_\beta\}=\{\bar Q^\alpha,\bar Q^\beta\}=0,\quad \{Q_\alpha,\bar Q^\beta\}=2\big((H-Y)\delta_{\alpha}^{\;\beta}+Z_{\alpha}^{\;\beta}\big),
\label{WSQM2}\end{equation}
where the tensor $\delta_{\alpha\beta}$ is replaced by $\delta_\alpha^{\;\beta}$,
and the charges $Z_{\alpha\beta}$ by $Z_\alpha^{\;\beta}$. Denote also
$$
\psi^2=\psi_\alpha\psi^\alpha=2\psi_1^-\psi_2^-,\quad
\bar\psi^2=\bar\psi^\alpha\bar\psi_\alpha=2\psi_2^+\psi_1^+.
$$

Then we have
$$
Y=\frac{c}{2}( \psi^2+\bar\psi^2),
\qquad Z_\alpha^{\;\beta}=c(\psi_\alpha\bar\psi^\beta+\psi^\beta\bar\psi_\alpha),
$$
which coincide with the representation given in \cite{Sweak,Sind}. As to the supercharges, we obtain
\begin{eqnarray*} &&
Q_\alpha=(\p+v(x))(\psi_\alpha - \bar\psi_\alpha)+B(x)(\psi_\alpha(1-\bar\psi^2) +(1+\psi^2)\bar\psi_\alpha),
\\ &&
\bar Q^\alpha=(-\p+v(x))(\psi^\alpha + \bar\psi^\alpha)+B(x)((1-\psi^2)\bar\psi^\alpha -\psi^\alpha(1+\bar\psi^2)).
\end{eqnarray*}
These expressions coincide with the $S_\alpha$ and $\bar S^\alpha$ supercharges given in
\cite{Sweak} (their terms cubic in fermions given in \cite{Sind} contain misprints).
Finally, we obtain the Hamiltonian
\begin{equation}
H=-\p^2+v^2(x)+B^2(x) +v'(x)(\psi^2+\bar\psi^2)
- B'(x) (\psi^2\bar\psi^2-2\psi_\alpha\bar\psi^\alpha+1).
\label{H_Smilga}\end{equation}
This Hamiltonian corresponds to the Lagrangian
$$
L=\big(\frac{dx}{dt}\big)^2+2i\bar\psi^\alpha\frac{d\psi_\alpha}{dt}-
v^2(x)-B^2(x) -v'(x)(\psi^2+\bar\psi^2)+ B'(x) (\psi^2\bar\psi^2-2\psi_\alpha\bar\psi^\alpha+1).
$$
Thus we have fully recovered the original starting point of Smilga in \cite{Sweak}.

The fermion operator projectors onto the $h_j$-subhamiltonian eigenfunctions have the following
cumbersome form
\begin{eqnarray} \label{P1} && \makebox[-2em]{}
P_j=\tfrac{1}{4}\left( 1+[\psi_1^+,\psi_1^-][\psi_2^+,\psi_2^-] +(-1)^j\tfrac{ 2Y}{c}
\right)
=\left(\begin{array}{cccc}
\delta_{j1} & 0 & 0 & 0 \\
0 & 0 & 0 & 0 \\
0 & 0 & 0 & 0 \\
0 & 0 & 0 & \delta_{j4} \\
\end{array}\right), \; j=1,4,
\\ \label{P2} && \makebox[-2em]{}
P_j=\tfrac{1}{4}\left( 1-[\psi_1^+,\psi_1^-][\psi_2^+,\psi_2^-] -(-1)^j\tfrac{ 2(Z_{12}+Z_{21})}{c}\right)
=\left(\begin{array}{cccc}
0 & 0 & 0 & 0 \\
0 & \delta_{j2} & 0 & 0 \\
0 & 0 & \delta_{j3} & 0 \\
0 & 0 & 0 & 0 \\
\end{array}\right), \; j=2,3.
\end{eqnarray}

Consider the structure of eigenfunctions of the Hamiltonian in terms of the Grassmann variables.
Since $\{\psi_\alpha,\bar\psi^\beta\}=\delta_\alpha^{\;\beta}$, one can replace in \eqref{H_Smilga}
fermionic operators by the Grassmann variables and their derivatives
\begin{equation}
\psi_\alpha \to \theta_\alpha,\quad
\bar\psi^\alpha \to \tfrac{\partial}{\partial\theta_\alpha}, \quad
\{\theta_\alpha,\theta_\beta\}=
\{\tfrac{\partial}{\partial\theta_\alpha},\tfrac{\partial}{\partial\theta_\beta}\}=0, \quad
\tfrac{\partial}{\partial\theta_\alpha} \theta_\beta=\delta_{\alpha\beta}.
\label{grassmann}\end{equation}
Then one can easily check that
$$
H\Phi(x)(1-\theta_1\theta_2)=h_1\Phi(x)(1-\theta_1\theta_2),\quad
H\Phi(x)(1+\theta_1\theta_2)=h_4\Phi(x)(1+\theta_1\theta_2),
$$
where $\Phi(x)$ is a scalar function (not a column). Thus, functions of the form $\Phi(x)(1\pm\theta_1\theta_2)$ describe the bosonic subspace of states. Analogously, functions of the form $\Phi(x)\theta_\alpha$
describe the fermionic states,
$$
H\Phi(x)\theta_\alpha=h_\beta\Phi(x)\theta_\alpha, \quad \alpha, \beta=1,2.
$$

Properties of the HSQM models essentially differ from those of the standard supersymmetric systems \cite{AIS}.
In particular, the double degeneracy may be starting only from the $(n+1)$-st energy level, since supercharges
kill in this case up to $n$ lowest energy eigenfunctions. As a result, the Witten index does not count
any more the difference between numbers of bosonic and  fermionic vacua. Instead, it starts to
depend on the chemical potential $\beta$ and characterizes the structure of zero modes of
supercharges.

Let us show that for $n=2$ HSQM models the Witten index essentially coincides with the superconformal
index \cite{Rom1,KMMR} for Smilga's one-dimensional weak supersymmetric model.
We definte the Witten index as
\begin{equation}
I_W=Tr \left((-1)^Fe^{-\beta H}\right), \quad F:=f^-f^+.
\label{Wind}\end{equation}
The $\Z_2$-grading operator $(-1)^F$ equals +1 for the upper element of the column of
HSQM Hamiltonian eigenfunctions, which is thus naturally identified with the bosonic subspace
of states. For the WSQM model \eqref{ham_wsqm}, the superconformal index
is defined for a distinguished pair  of supercharges, which we choose to be $Q_1^\pm$:
\begin{equation}
I_{SCI}=Tr \left((-1)^Fe^{-\gamma\{Q_1^-,Q_1^+\}}e^{-\beta H}\right), \quad F=\tfrac{1}{c}Y+1.
\label{SCI}\end{equation}
Here the $\Z_2$-grading operator $(-1)^F$ has eigenvalues +1 for the highest and lowest terms
of the column of Hamiltonian eigenfunctions forming the bosonic subspace of states.
The two middle terms in the latter column form the fermionic sector since $(-1)^F$ becomes
equal to -1 for them.
Dependence on the chemical potential $\gamma$ is actually absent for the same reason
as the Witten index does not depend on $\beta$ in the standard situation. However,
dependence on $\beta$ in \eqref{SCI} is quite nontrivial, since it characterizes
the structure of BPS states killed by the supercharges $Q_1^\pm$.

In order to classify admissible structures of the lowest energy levels and determine the form of these indices
one should list possible situations when zero modes of the operators $A^\pm_\alpha$
$$
A^\pm_\alpha \phi_\alpha^\pm(x)=0, \quad \phi_\alpha^\pm(x)=\exp\big(\pm \int^xv_\alpha(y)dy\big),
$$
 define physical states and when they don't. We drop arbitrary constant multiplicative factors since
 it is sufficient  to see if these functions belong to $\mathrm{L}^2(\R)$.
If normalizable, they represent ground states and thus should not have zeroes.
To find particular examples of $\phi_\alpha^\pm(x)$ with different $x\to \pm \infty$ asymptotic
behaviour, one can use the freedom in the function $v(x)$ or $B(x)$.

Figure 1 describes possible energy spectra for all the types of models PSQM, HSQM, and WSQM at
once. Full pictures correspond to HSQM. If one deletes one of the middle towers of states, then
the pictures reduce to the PSQM Hamiltonians spectra almost exactly as they were
described in  \cite{RS}. If both middle towers are removed, then they describe the $n=2$ HSQM  case.

\begin{figure}
\centering
\includegraphics[width=120mm]{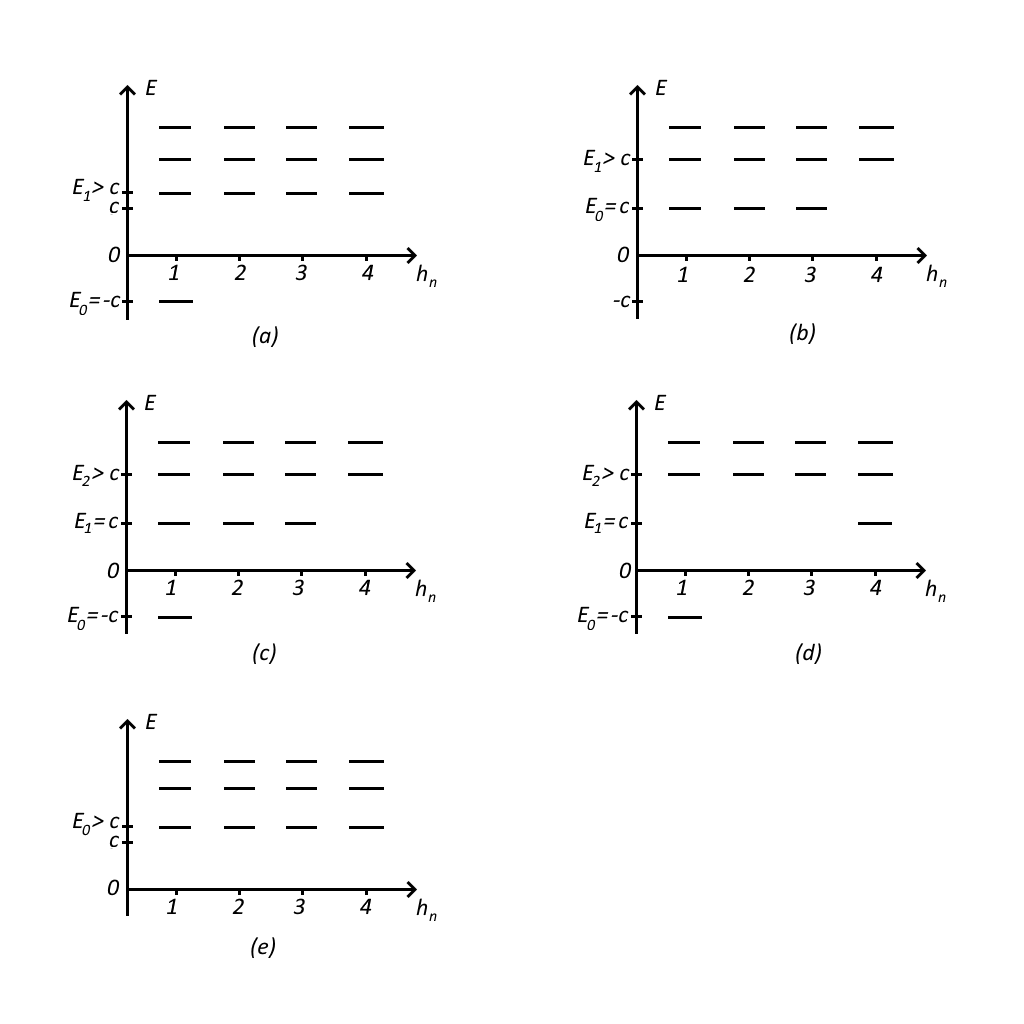}
\vspace{-11mm}
\caption{Forms of the spectra for the WSQM Hamiltonian.}
\label{figKP}
\end{figure}

a) Suppose that $\phi_1^-(x)$ represents the ground state of the $h_1$ subhamiltonian. Then the function
$\phi_1^+(x)$ cannot be normalizable. Let $\phi_2^\pm(x)$ be unphysical (not normalizable).
The constant $c$ may be positive or negative. In Fig. 1a we depicted this case situation for $c>0$.
Then the vacuum states of $H_{WSQM}$ and $H_{HSQM}$ with the  energy $E_0=-c$ are
$$
\psi_{E_0}(x)=(\phi_1^-(x), 0,0,0)^t=\phi_1^-(x)(1-\theta_1\theta_2)\quad \textrm{and}\quad
\psi_{E_0}(x)=(\phi_1^-(x), 0)^t,
$$
respectively.
They satisfy the conditions $Q_1^\pm\psi_{E_0}(x)=0$ (WSQM) and $Q^\pm\psi_{E_0}(x)=0$ (HSQM) and yield
$$
I_W=I_{SCI}=e^{\beta c}.
$$
So, supersymmetry is not broken in this case.
The second eigenvalues of these Hamitonians satisfy the constraint $E_1>|c|$.

An example of such a lowest levels structure is given by the following functions
$$
v_1(x)=\alpha\sinh\beta x-\tfrac12 \beta, \quad
v_2(x)= \alpha\cosh\beta x+\tfrac12 \beta, \quad \alpha>0,
$$
which yields $c=-\tfrac12 \alpha^2<0$.

b) Suppose now that only $\phi_2^-(x)$ is a normalizable function.
This case is realizable only for $c>0$. Indeed, eigenvalues of $h_1$ must be bigger than $-c$,
whereas isospectral to it subhamiltonians  $h_2$ and $h_3$ have the ground state energy $E_0=c$.
This situation is depicted in Fig. 1b. One has a triply degenerate ground state:
$$
\psi_{E_0}(x)\propto (0,0,\phi_2^-(x),0)^t,\quad (0,\phi_2^-(x),0,0)^t, \quad (A_1^+\phi_2^-(x),0,0,0)^t.
$$
Still, there is only one their combination satisfying the BPS condition $Q_1^\pm\psi_{BPS}(x)=0$,
which is a fermionic state
$$
\psi_{BPS}(x)=(0,\phi_2^-(x), - \phi_2^-(x),0)^t=\phi_2^-(x)\theta_2.
$$
Only this state gives
non-zero contribution to $I_{SCI}$. The corresponding HSQM model has $\psi_{E_0}(x)=(A_1^+\phi_2^-(x),0)^t$
with $E_0=c$ and $Q^\pm \psi_{E_0}(x)=0$. So, supersymmetry is not broken in both cases.
This yields the indices
$$
I_W=-I_{SCI}=e^{-\beta c}.
$$
The second energy level $E_1>c$ remains unknown in general.

An example of such a configuration is obtained just by flipping the $\cosh\beta x$ and $\sinh\beta x$
functions in the previous example
$$
v_1(x)= \alpha\cosh\beta x-\tfrac12 \beta,
\quad
v_2(x)=\alpha\sinh\beta x+\tfrac12 \beta,  \quad \alpha>0,
$$
which yields $c=\tfrac12 \alpha^2>0$.

c) The third possibility corresponds to the case when both $\phi_\alpha^-(x)$ represent physical states.
For $c>0$ this situation is depicted in Fig. 1c. The WSQM ground state is
$$
\psi_{E_0}(x)=(\phi_1^-(x), 0,0,0)^t=\phi_1^-(x)(1-\theta_1\theta_2), \quad E_0=-c,
$$
and this is a BPS state, $Q^\pm_1\psi_{E_0}(x)=0$. The first excited level has the energy $E_1=c$
and it is triply degenerate: $\psi_{E_1}(x)\propto (0,0,\phi_2^-(x),0)^t,\, (0,\phi_2^-(x),0,0)^t,\, (A_1^+\phi_2^-(x),0,0,0)^t$. Only one their particular fermionic combination forms a BPS state,
$$
\psi_{BPS}(x)=(0,\phi_2^-(x), - \phi_2^-(x),0)^t=\phi_2^-(x)\theta_2,\quad
Q_1^\pm\psi_{BPS}(x)=0.
$$

In the  HSQM case, one has the ground state $\psi_{E_0}(x)=(\phi_1^-(x),0)^t$ with $E_0=-c$
and the first excited level $\psi_{E_1}(x)=(A_1^+\phi_2^-(x),0))^t$ with $E_1=c$, which
both are annihilated by the supercharges $Q^\pm \psi_{E_{0,1}}(x)=0$. There are thus
two nonzero contributions both to $I_W$ and $I_{SCI}$:
$$
I_W=e^{\beta c}+e^{-\beta c}, \quad I_{SCI}=e^{\beta c}-e^{-\beta c}.
$$
The supersymmetry is not broken and the next excited level energy $E_2>c$ remains unknown.

The simplest example of such a situation is given by the harmonic oscillator
$$
v_1(x)=v_2(x)=\omega x, \quad c=\omega>0.
$$

d) Suppose now that $\phi_1^-(x)$ and $\phi_2^+(x)$ represent physical states. Then both $c>0$ and $c<0$
cases are possible. For $c>0$ the structure of spectrum is depicted in Fig. 1d.
The WSQM ground state is
$$
\psi_{E_0}(x)=(\phi_1^-(x),0,0,0)^t=\phi_1^-(x)(1-\theta_1\theta_2), \quad E_0=-c
$$
and the first excited state
$$
\psi_{E_1}(x)=(0,0,0,\phi_2^+(x))^t=\phi_2^+(x)(1+\theta_1\theta_2),\quad E_1=c.
$$
Both of them represent
BPS states, $Q_1^\pm\psi_{E_0}(x)=Q_1^\pm\psi_{E_1}(x)=0$. For HSQM one has the lowest energy eigenfunctions
obtained from the described column of WSQM states by deletion of two middle terms.
Therefore there are two nontrivial contributions to both indices $I_W$ and $I_{SCI}$ entering them
with different signs:
$$
I_W=e^{\beta c}-e^{-\beta c}, \quad I_{SCI}=e^{\beta c}+e^{-\beta c}.
$$
The supersymmetry is not broken and the eigenvalue $E_2>c$ remains unknown.

An example of such a configuration is
$$
v_1(x)=\alpha\tanh (\alpha-\beta)x,\quad
v_2(x)= \beta\tanh (\alpha-\beta)x,\quad \beta< 0< \alpha,\quad c=\tfrac12(\alpha^2-\beta^2).
$$

e) The final configuration, depicted in Fig. 1e, corresponds to the situation when all $\phi_\alpha^\pm(x)$
are not normalizable. The supersymmetry is broken since the WSQM model has no BPS states and for the HSQM model
supercharges do not kill vacua. The WSQM and HSQM ground states are four-fold and two-fold degenerate, respectively, and their energy is not known $E_0 > |c|$. As a result,
$$
I_W=I_{SCI}=0.
$$

An example of such a configuration is given by the choice of $g(x)$ in \eqref{g} as the Gaussian function
$g(x)=\exp(-\alpha x^2)$, $\alpha>0$, which yields $B(x)=-\alpha x$ and
$$
v_j(x)=(-1)^j\alpha x+c e^{\alpha x^2}\int_{x_0}^x e^{-\alpha y^2}dy,\quad j=1,2,
$$
with an arbitrary parameter $c\neq 0$.

Other formally possible options are as follows. For example,
if only $\phi_2^+(x)$ is normalizable, then, $\psi_{E_0}(x)\propto (0,0,0,\phi_2^+(x))^t$
with $E_0=c$. By the transformations $v_1(x)\leftrightarrow -v_2(x), \, c\to -c$,
this case is mapped to the model a).
If both $\phi_\alpha^+(x)\in\mathrm{L}^2(\R)$, then by the same transformations
one comes to the case c). The situation when $\phi_1^+(x),\, \phi_2^-(x)\in\mathrm{L}^2(\R)$,
is not possible.
For $c>0$, the function $\phi_1^+(x)$ cannot be normalizable alone. This corresponds to
the subhamiltonians $h_2$ and $h_3$ having the ground state energy $E_0=-c$, whereas the
isospectral to them hamiltonian $h_4$ should have eigenvalues bigger than $c$.

Our analysis shows that the Witten index $I_W$ for $n=2$ HSQM models introduced in \cite{AIS}
 is actually the disguised
superconformal index $I_{SCI}$ for WSQM systems, since they both describe the structure
of lowest energy states almost in an identical way. The difference in signs of terms in $I_{SCI}$ and $I_W$
comes from a change of prescription whether the corresponding terms come
from the fermionic or bosonic states. One can also note that the spectra described in figures 1a) and 1e)
coincide with the ones of the canonical $\mathcal{N}=2$ SQM Hamiltonian
(with exact and spontaneously broken supersymmetry, respectively) after shifting it by a constant.
The equivalence of $n=2$ HSQM and WSQM models was anticipated in \cite{Sweak}
and recently Smilga computed in \cite{Sind} the index $I_{SCI}$ for a one-dimensional
model corresponding to the situation of Fig. 1c. In the present paper we have
given full classification of all possible one-dimensional cases, computed $I_{SCI}$
for them and gave a detailed comparison of arising models with those considered in
the old works \cite{RS} and \cite{AIS}. It would be interesting to understand what kind of
WSQM models are hidden behind the general $n>2$ HSQM systems. Some of them were
considered in \cite{S2022}, but the variety of admissible systems should be much richer,
as it is shown here already for the $n=2$ case,

As a last remark, let us show that the weak supersymmetric model under consideration is a
special interpretation of a special case of the $4\times 4$ matrix SQM model considered in \cite{AISV}.
Define unitary matrices
$$
\mu_l=\left(\begin{array}{cccc}
0 & 1 & 0 & 0 \\
0 & 0 & 1 & 0 \\
1 & 0 & 0 & 0 \\
0 & 0 & 0 & 1 \\
\end{array}\right), \quad
\mu_r=\left(\begin{array}{cccc}
0 & 0 & 1 & 0 \\
1 & 0 & 0 & 0 \\
0 & 1 & 0 & 0 \\
0 & 0 & 0 & 1 \\
\end{array}\right), \quad
V=\frac{1}{\sqrt2}\left(\begin{array}{cccc}
-1 & -1 & 0 & 0 \\
1 & -1 & 0 & 0 \\
0 &0 & \sqrt2 & 0 \\
0 & 0 & 0 & \sqrt2 \\
\end{array}\right),
$$
where $\mu_l=\mu_r^{-1}$ and $V^\dag=V^{-1}$.
The hermitian supercharge
$$
Q_2^{(1)}=\mu_l(Q_2^+ + Q_2^-)\mu_r =
\left(\begin{array}{cccc}
0 & 0 & -A_1^- & A_2^+ \\
0 & 0 & A_1^- & A_2^+ \\
-A_1^+ & A_1^+ & 0 & 0 \\
A_2^- & A_2^- & 0 & 0 \\
\end{array}\right)
$$
coincides in structure with the general $N=1$ SQM model supercharge considered in \cite{AISV}.
Performing another unitary transformation, we get a maximally simplified form of the corresponding model
$$
\mathcal{Q}_2^{(1)}=V^{-1}Q_2^{(1)}V=\sqrt2 (q_2^+ +q_2^-), \quad \{q_2^+,q_2^-\}= \mathcal{H},
\quad (q_2^\pm)^2=[\mathcal{H}, q_2^\pm]=0,
$$
where
$$
q_2^+= \left(\begin{array}{cc}
0 & -\p +\hat v(x)\\
0 & 0 \\
\end{array}\right),
\quad q_2^-=\left(\begin{array}{cc}
0 & 0 \\
\p +\hat v(x) & 0 \\
\end{array}\right),
\quad \hat v(x)=\left(\begin{array}{cc}
-v_1(x) & 0\\
0 & v_2(x) \\
\end{array}\right).
$$
In \cite{AISV} the matrix superpotential $\hat v(x)$ had the general $2\times 2$ form with three
 arbitrary functions $v_j(x),\, j=0,1,2$, where $v_0$ (denoted as $W_0$ in \cite{AISV}) defined
 antidiagonal matrix elements.
Therefore, in general one had only $N=1$ supersymmetric algebra $su(1|1)$.
In the present case we have $v_0=0$ which opens up
new symmetries of the model leading to the centrally extended $su(2|1)$ algebra, as described above.
The Hamiltonian $\mathcal{H}$ does not coincide with the Hamiltonian $H$ \eqref{ham_wsqm},
$$
\mathcal{H}=-\p^2+\left(\begin{array}{cc}
\hat v^2(x)-\hat v'(x) & 0\\
0 & \hat v^2(x)+\hat v'(x) \\
\end{array}\right)=\textup{diag}(A_1^-A_1^+, A_2^+A_2^-,A_1^+A_1^-,A_2^-A_
2^+),
$$
i.e. the physical content of these two described interpretations of supersymmetry is different.
Note also a sharp difference in the anticommutators of supercharges: $\{q^+,q^-\}$
is a diagonal matrix, whereas $\{Q_1^+,Q_1^-\}$ and $\{Q_2^+,Q_2^-\}$ are not diagonal.

Defining other hermitian supercharges
\begin{eqnarray*} && \makebox[6em]{}
\mathcal{Q}_2^{(2)}=-\textup{i} V^{-1}\mu_l(Q_2^+ - Q_2^-)\mu_rV,
\\ &&
\mathcal{Q}_1^{(1)}=V^{-1}\mu_l(Q_1^+ + Q_1^-)\mu_rV,\quad
\mathcal{Q}_1^{(2)}=-\textup{i} V^{-1}\mu_l(Q_1^+ - Q_1^-)\mu_rV,
\end{eqnarray*}
we come to the algebra
$$
\{\mathcal{Q}_\alpha^{(i)}, \mathcal{Q}_\beta^{(j)}\}
=4\left( (H_p-Y)\delta_{ij}\delta_{\alpha\beta}+Z^{ij}_{\alpha\beta}\right),
\quad [H_p,\mathcal{Q}_\alpha^{(i)}]=[H_p,Y]=[H_p,Z^{ij}_{\alpha\beta}]=0,
$$
where $H_p=\textrm{diag}(h_2,h_3,h_1,h_4)$ (it differs from \eqref{ham_wsqm} only by a
permutation of $h_j$-subhamiltonians) and the charges $Y, Z^{ij}_{\alpha\beta}$
are defined by the following $4\times 4$ matrices
\begin{eqnarray*} && \makebox[1.5em]{}
Y=-c\left(\begin{array}{cc}
0 & 0\\
0 & \sigma_3 \\
\end{array}\right), \quad
Z_{11}^{ij}=-Z_{22}^{ij}= -c\delta_{ij}\Sigma_3,
\\ &&
Z_{12}^{ii}=-c\Sigma_1, \quad
Z_{12}^{12}=Z_{21}^{12}=c\Sigma_2, \quad
\Sigma_k= \left(\begin{array}{cc}
\sigma_k & 0\\
0 & 0 \\
\end{array}\right),
\end{eqnarray*}
with $\sigma_k$ being the ordinary Pauli matrices.

To conclude, although the author was computing superconformal indices in four-dimensional field theories
for a rather long time \cite{SV,SV2}, their connection with the early works  \cite{RS,AISV,AIS}
was understood only after knowing the paper \cite{Sweak}. It was surprising to learn that
there exists a direct road from the rather elementary considerations in modified supersymmetric quantum
mechanical models to complicated modern mathematical constructions in the theory of special functions \cite{rev}.

\smallskip

{\bf Acknowledgements.}
Some results of the present paper were announced in the Proceedings of the conference dedicated
to the memory of V. A. Rubakov \cite{PoS}. The whole study was performed in a wish to stress
an importance of the joint work with Valery \cite{RS} for author's eventual landing on
the new class of special functions describing superconformal indices in $4d$ field theories.
The author is indebted to A. V. Smilga for discussions on the weak supersymmetry and related
quantum mechanical models.
This study has been partially supported by the Russian Science Foundation (grant 24-21-00466).

\end{document}